\documentclass[twocolumn,floatfix,preprintnumbers]{revtex4}
\usepackage{graphics,amssymb,amsmath,epsfig,color}
\usepackage{graphicx}
\usepackage{subfigure}

\usepackage[T1]{fontenc}
\usepackage{ifsym}
\usepackage{wasysym}
\usepackage{babel}
\usepackage[normalem]{ulem}
\usepackage{soul}
\usepackage{natbib}
\usepackage{xr}
\usepackage{hyperref}
\begin{document}
	\title{Experimental observation of the origin and structure of elasto-inertial turbulence}
	\author{George H. Choueiri$^{1,2}$,  Jose M. Lopez$^1$, Atul Varshney$^1$, Sarath Sankar$^1$, and Bj\"orn Hof$^1$}                     
	\affiliation{$^1$Institute of Science and Technology Austria, 3400 Klosterneuburg, Austria\\
	$^2$MIME Department, University of Toledo, Toledo, Ohio, United States}

	\begin{abstract}Turbulence generally arises in shear flows if velocities and hence inertial forces are sufficiently large. In striking contrast,  viscoelastic fluids can exhibit disordered motion even at vanishing inertia. Intermediate between these cases, a novel state of chaotic motion, `elasto-inertial turbulence' (EIT), has been observed in a narrow Reynolds number interval. We here determine the origin  of EIT in experiments and show that characteristic EIT structures can be detected across an unexpectedly wide range of parameters. Close to onset a pattern of chevron shaped streaks emerges in excellent agreement with linear theory. However, the instability can be traced to far lower Reynolds numbers than permitted by theory. For increasing inertia, a secondary instability gives rise to a wall mode composed of inclined near wall streaks and shear layers. This mode persists to what is known as the `maximum drag reduction limit' and overall EIT is found to dominate viscoelastic flows across more than three orders of magnitude in Reynolds number.
		\end{abstract}
	\maketitle
	%
	Many fluids in nature and applications, such as paints, polymer melts or saliva have viscous as well as elastic properties and their flow dynamics fundamentally differs from that of  Newtonian fluids. A standard example of such viscoelastic fluids are solutions of long chain polymers and here surprisingly even very dilute solutions show a drastic suppression of turbulence and significantly lower drag levels \cite{Pro08,White_Mungal08}; a phenomenon commonly exploited in pipeline flows to save pumping costs. In seeming contradiction to this stabilizing effect are observations at much lower Reynolds numbers ($Re$, the ratio of inertial to viscous forces), where polymers have the exact opposite effect; they initiate fluctuations and increase the flow's drag. The resulting chaotic motion was first detected in a narrow Reynolds number interval, $1000 \lessapprox Re \lessapprox 2000$, just below the onset of ordinary turbulence~\cite{RaTa60,LiWie70} and interpreted as a form of early turbulence. However, it was later shown~\cite{Samanta13} that the corresponding elasto-inertial instability can be traced into the polymer drag reduction regime at larger $Re$. The suggestion of a possible connection between these two seemingly opposing effects has sparked much recent interest  in the phenomenon of elasto-inertial turbulence (EIT) \cite{dubief13,LoChHo19,ShMcMuWaMkeGra19,garg2018,Cho18,chandra_onset_2018,page_exact_2020,chandra_early_2020}. 
	
	It has additionally been speculated that EIT may be connected to purely `elastic turbulence'; a fluctuating state driven by a linear elastic instability in the inertialess limit~\cite{groisman2000elastic}. This instability requires curved streamlines~\cite{groisman2000elastic,larson_shaqfeh_muller_1990,shaqfeh1996purely} and is hence not to be expected in flows through smooth straight pipes. However, recent findings tend to suggest that a similar instability mechanism may also occur in planar shear flows following an array of strong perturbations~\cite{arratia13,Qin08}, leading to pure elastic turbulence through a subcritical transition scenario.

	Although EIT has first been observed in pipe flow experiments ~\cite{RaTa60,LiWie70}, information on the structure and nature of the resulting state is almost exclusively based on simulations using polymer models. Such simulations and theoretical considerations have suggested a range of  possible transition scenarios. In direct numerical simulations (employing the FENE-P model) the characteristic features of EIT include near wall vortical structures oriented perpendicular to the mean flow direction (i.e. spanwise direction) and elongated sheets of constant polymer stretch inclined with respect to the wall. In these simulations, the transition leading to this state is nonlinear, (i.e. subcritical) and requires perturbations of finite amplitude~\cite{Samanta13,dubief13,LoChHo19}. In another study the aforementioned spanwise vortical structures were suggested to be linked to the well known Tollmien-Schlichting (TS) instability that occurs in channel flow of Newtonian fluids at substantially larger Reynolds numbers. Again here the transition would be subcritical, however linked to TS waves~\cite{ShMcMuWaMkeGra19}. Yet other studies reported a linear instability that gives rise to chevron shaped streaks~\cite{garg2018}. The latter proposed that this supercritical transition may be the starting point of a sequence of instabilities that eventually lead to EIT.

	In the present study we visualize the onset of EIT in experiments and show that the flow pattern is in excellent agreement with the unstable mode predicted by linear stability analysis~\cite{garg2018}. However in experiments fluctuations are already present close to onset suggesting that nonlinear effects cannot be neglected. Moreover, for increasing shear rates the instability can be pushed to $Re$ an order of magnitude below the parameter regime predicted by linear analysis. For increasing $Re$ on the other hand the dominant flow structures adjust from a centre to a wall mode and fluctuation levels strongly increase. The resulting three dimensional EIT flow pattern persists to the so called `maximum drag reduction' (MDR) regime at much larger $Re$. Structural features of EIT can hence be detected across more than three decades in $Re$.

\begin{figure}[t!]
		\includegraphics[width=\linewidth]{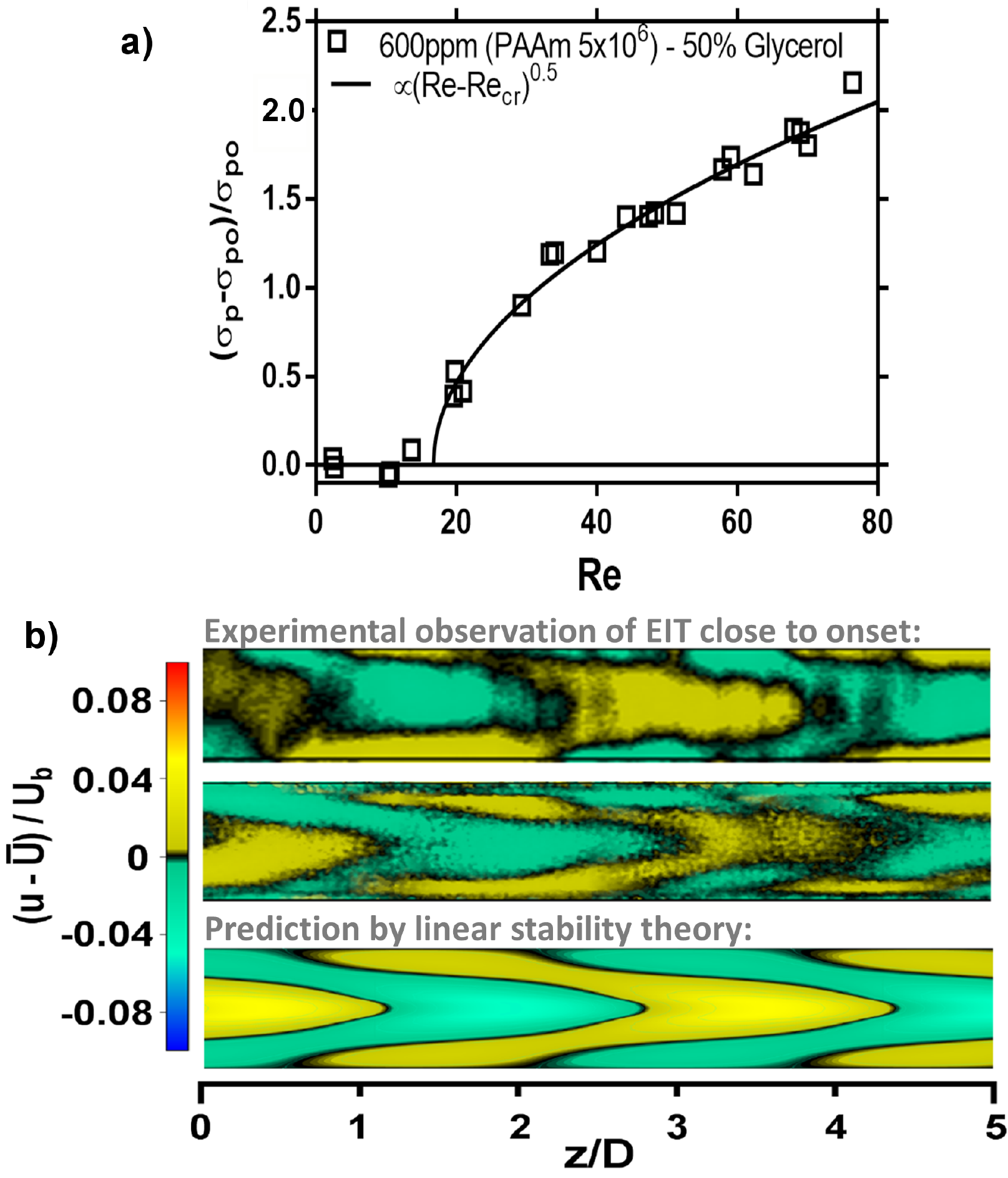}  
		\caption{\label{fig:instability_onset} \textbf{Fluctuations level and flow structure near the onset of elasto-inertial instability}. Figure (a) shows the evolution of the pressure fluctuations amplitude with increasing $Re$ close to the instability threshold for experiments using 600 ppm of PAAm dissolved in a 50\% water glycerol mixture. The symbol $\sigma_{p}$ denotes the standard deviation of the pressure fluctuations, $\sqrt{<p^2>}$, whereas $\sigma_{p_{0}}$ indicates the standard deviation of the background noise level for the pressure sensor, $\sqrt{<p_{0}^2>}$. Figure (b) illustrates the flow's structural composition at Reynolds numbers near transition. 
			The upper and middle panels show streamwise velocity fluctuations obtained from PIV measurements in a longitudinal cross section. The upper panel shows flow structures at $Re \approx 5$ and corresponds to an experiment using a 66\% glycerol concentration ($\beta = 0.56$, $W_i = 96$), whereas the middle panel corresponds to $Re \approx 100$ in an experiment using a 50\% glycerol concentration ($\beta = 0.56$, $W_i = 203$). The lower panel shows the most unstable mode in the linear stability analysis; the solution plotted was calculated for $Re = 100$, $W_i = 60$, $\beta = 0.9$, $n=0$ and $k=1$ (see SI for the definitions of these parameters). Flow direction is from right to left.}
	\end{figure}

	Experiments were initially performed using a $50\%$ water glycerol mixture as solvent and dissolving 600 ppm polyacrylamide with a molecular weight of $5 \times 10^6$ Da. The resulting  solution has a viscosity of $\approx 10.2$ times that of water. The standard deviation of the pressure fluctuations recorded for increasing Reynolds numbers is plotted in  Fig.~\ref{fig:instability_onset}(a). The fluctuation level is initially zero (when subtracting the sensor's background noise level), meaning that the flow is laminar, but begins to grow at $Re \approx 18$, as the elasto-inertial instability sets in. After the onset of instability, the fluctuation amplitude grows continuously with increasing $Re$; approximately in proportion to the square root of $Re$ as indicated by the solid line. Nevertheless, given the small amplitudes and experimental uncertainties other scaling relations cannot be ruled out.

	Structural information is obtained from the velocity fields recorded in the pipe's central plane using particle image velocimetry (PIV). The instantaneous snapshots are assembled by applying Taylor's frozen-flow hypothesis and the resulting flow structure is shown in the mid panel of  Fig.~\ref{fig:instability_onset}(b) for $Re \approx 100$. For visualization purposes the average cross-sectional velocity profile $\bar{U}$ is subtracted from the data and areas with velocities lower (higher) than the mean profile are shown in green (yellow). These low and high speed streaks alternate in the streamwise direction and show a tendency to form a chevron type pattern. The streak amplitudes are less than $5\%$ of $U_b$ (the mean velocity) and therefore lower than streak amplitudes in ordinary turbulent flow. 
	To compare these flow patterns with the unstable mode predicted by the linear stability theory we repeated the analysis in~\cite{garg2018}, however using a different methodology (see SI for details). The obtained results are in excellent agreement with those in~\cite{garg2018}. The least stable mode for $Re=100$ is shown in the bottom panel of  Fig.~\ref{fig:instability_onset}(b). As seen, here also a chevron type pattern consisting of alternating low and high speed streaks is observed. Hence, the least stable mode can be detected in experiments suggesting that the elasto inertial instability mechanism described in \cite{garg2018} is indeed central to the onset of EIT. However, while the stability analysis predicts a perfectly regular structure
	(resulting from a super-critical Hopf bifurcation), in our case the structure is not singly periodic but fluctuations appear across a range of frequencies suggesting weakly chaotic flow. Attempts to resolve the flow field closer to onset of instability for the same fluid were unsuccessful due to the lower signal to noise ratio. 
	
	In order to probe if the elasto inertial instability persists to even lower $Re$ additional experiments were carried out for a $66\%$ glycerol water solution again adding $600$ ppm of PAAm. Due to the increased solvent viscosity (approximately three times higher) the shear rates at a given $Re$ increase by the same factor compared to the $50\%$ solution. At the same time and as reported in~\cite{Samanta13}, for a given polymer type and concentration, the onset of the elasto-inertial instability is dictated by the shear rate and hence it is expected to set in at lower $Re$. Indeed, for the $66\%$ glycerol water mixture, fluctuations were detected at $Re$ as low as five whereas at slightly lower $Re\sim3$ the flow was laminar (not shown). As shown in the top panel of  Fig.~\ref{fig:instability_onset}(b), at $Re=5$ the flow again consists of alternating high and low speed streaks arranged in a chevron pattern. It is noteworthy that according to the linear analysis~\cite{garg2018} the instability can only be continued to $Re\sim80$ or so, this however does not rule out the possibility that the same mode may occur subcritically at even lower $Re$. Moreover the flow pattern observed in our experiments at $Re=5$ is again not perfectly periodic (unlike predicted by linear theory) but still weakly chaotic. Both these observations are consistent with a subcritical scenario where the minimum amplitude threshold to trigger the instability, albeit finite, is low compared to the disturbance levels induced by typical experimental imperfections. In such a situation EIT would arise automatically even though in principle the laminar state is linearly stable. The irregularities of the flow pattern observed are an indication that the state has undergone further bifurcations and the resulting flow is chaotic and three dimensional in nature. 
	
	\begin{figure}[t!]
		\centering
		\includegraphics[scale = 0.65]{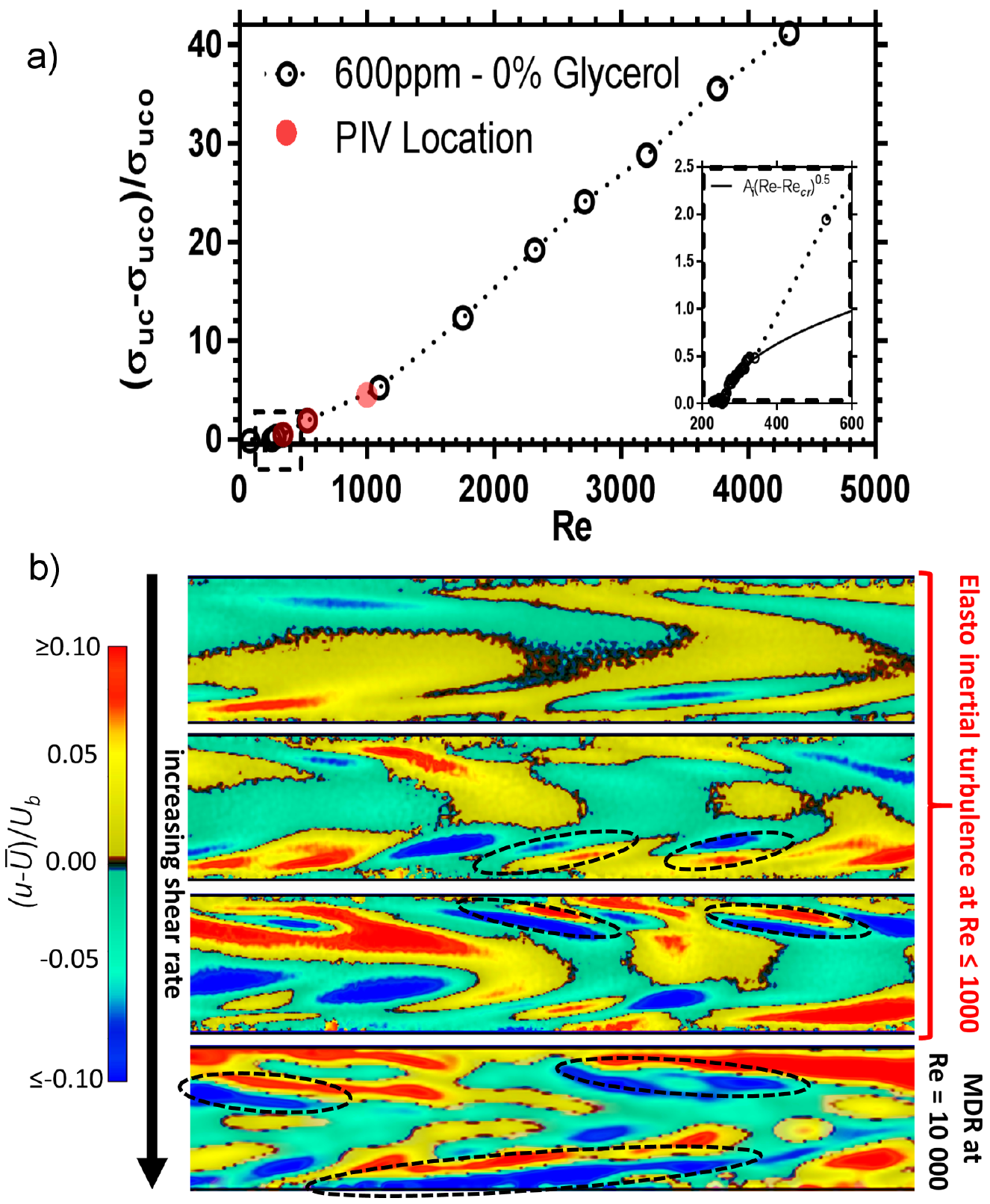}
		\caption{\label{fig:far_onset} \textbf{Flow development far from the instability onset.} (a) Variation of the streamwise velocity fluctuations at the pipe center with increasing $Re$ in experiments using 600 ppm of PAAm dissolved in water. The symbols $\sigma_{u_c}$ and  $\sigma_{u_{co}}$ denote the standard deviation of
			streamwise velocity fluctuations at the pipe centerline,$\sqrt{<u_c^2>}$, and the standard deviation of the background noise level for the LDV system, $\sqrt{<u_{co}^2>}$, respectively. The inset shows the deviation from the square root scaling. The red dots indicate the locations where the flow structures shown in panel (b) were measured. (b) PIV visualizations of streamwise velocity fluctuations. From top to bottom: $Re =$ 300, $Re =$ 500, $Re =$ 1000 and $Re=$ 10,000. With increasing shear large amplitude streaks (red, blue) arise in the near wall region. Streaks become more elongated with $Re$ and are separated by inclined shear layers (indicated by the dashed contours). }
	\end{figure}
	
\begin{figure}[t!]
		\centering
		\includegraphics[width=1\linewidth]{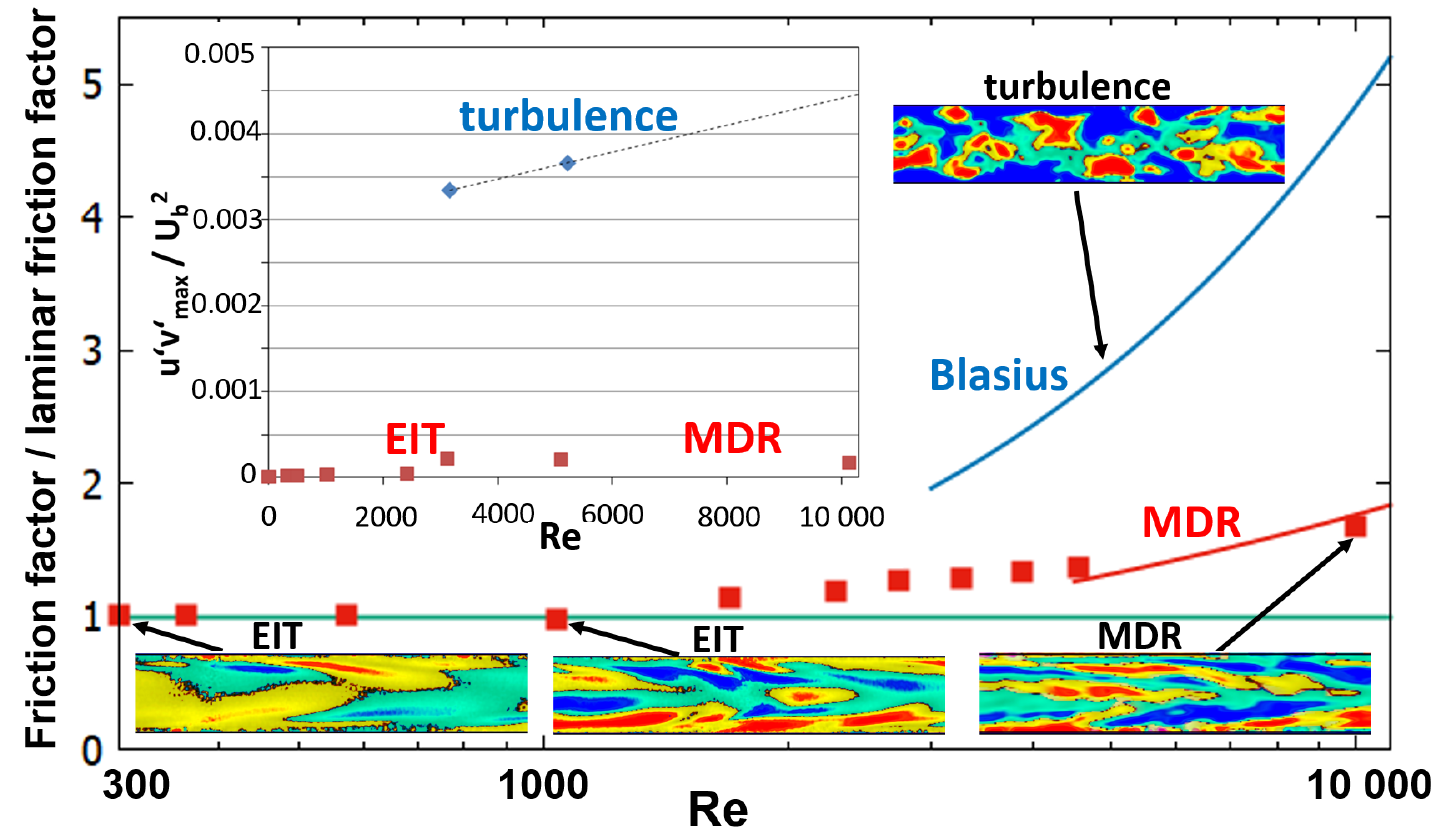}         
		\caption{\label{fig:MDR_friction} \textbf{Evolution of friction factors with increasing $Re$.} At low $Re$ the friction factors of EIT (red points) are indistinguishable from the laminar friction values (green line). With increasing $Re$ the friction factors visibly exceed the laminar level and continuously approach the MDR asymptote (red curve). The inset shows Reynolds stresses normalized by the square of the bulk flow velocity for Newtonian turbulence. EIT and MDR are at the same level, whereas turbulent Reynolds stresses are an order of magnitude larger.}
	\end{figure}

	We next investigate the further development of the flow pattern with increasing inertia (higher $Re$). In order to reach larger $Re$ the solvent was changed to water, again dissolving 600 ppm of PAAm. Owing to the reduced solvent viscosity, the onset of instability shifts to larger $Re$ ($\gtrsim$ 200). Also for the 600ppm PAAm solution in water the transition appears to be continuous (inset of  Fig.~\ref{fig:far_onset}(a)). With increasing $Re$ the fluctuation level does not saturate but instead begins to increase faster and subsequently the scaling becomes closer to linear (Fig.~\ref{fig:far_onset}(a)). At the lowest $Re$ ($=300$) where PIV measurements were carried out, the flow pattern still bears some resemblance to the chevron pattern (yellow and green isolevels shown in the top panel of  Fig.~\ref{fig:far_onset}(b)), however in the near wall region higher amplitude streaks (red and blue isolevels) have appeared. With a further increase in $Re$ and as the fluctuation level of the flow begins to increase more steeply, these near wall streaks become the predominant structure and the chevron mode in the central region of the pipe disappears (see second panel in  Fig.~\ref{fig:far_onset}(b)). Note that this mode change is equally found in the higher viscosity solvents (50\% and 66\% glycerol concentrations) for $Re$ sufficiently larger than those shown in  Fig.~\ref{fig:instability_onset}(b). With increasing shear levels low and high speed streaks often appear in pairs that are approximately parallel, signifying strong shear layers at the respective interface (see dashed black contours in Fig. \ref{fig:far_onset}(b)). Shear layers just like streaks are inclined with respect to the main flow direction and become more elongated with increasing $Re$, often extending over multiple pipe diameters. We interpret this mode change as a secondary instability. As $Re$ is further increased to $Re=1000$ (third panel in  Fig.~\ref{fig:far_onset}(b)) there is surprisingly little change in the overall flow structure and the wall mode continues to dominate the dynamics. Even for a tenfold increase, to $Re=10,000$ (bottom panel in Fig.~\ref{fig:far_onset}(b)) and hence a value that is well into the classical polymer drag reduction regime, the wall mode persists and the flow's structural composition closely resembles that of EIT at $Re=1000$, while it is clearly distinct from Newtonian turbulence (Fig. \ref{fig:MDR_friction} inset).

	In addition to velocity measurements, the pressure drop was recorded as $Re = 10,000$ was approached, which allows to directly determine fluid drag. The corresponding friction factors relative to the laminar level are shown in Fig. \ref{fig:MDR_friction}. At low Reynolds numbers friction factors of EIT (red
	points) only marginally exceed the laminar friction. With increasing Re deviations become notable and the friction values smoothly approach what is known as the maximum drag reduction
	(MDR) or Virk's asymptote (red line).  Regardless of the type of polymer, solvent, and relative concentration, Virk's asymptote sets a universal limit to the amount of drag reduction obtainable. Traditionally, MDR has been proposed as a residual, minimal level of ordinary turbulence and a relation to the edge state of Newtonian turbulence has been suggested~\cite{Pro08,White_Mungal08,XiGra10,XiGra12}. An interpretation that does not readily explain why polymers cannot reduce the drag beyond this level (reduction beyond MDR can be achieved in a narrow parameter regime only \cite{Cho18}, but not at high $Re$). As first suggested in~\cite{Samanta13}, the MDR scaling may instead be caused by the EIT instability, i.e. although polymers can largely suppress ordinary Newtonian type turbulence, eventually when shear levels are sufficiently large the elasto-inertial instability necessarily must arise inhibiting laminarisation.

	It is noteworthy that in the present study as the high inertia regime is approached the MDR friction scaling monotonically arises from low Reynolds number EIT. Structurally MDR and EIT are equally composed of elongated inclined streaks and shear layers. In contrast streaks in Newtonian turbulence are  shorter and less coherent (see panels in Fig. \ref{fig:MDR_friction}). In addition to the structural composition and the skin friction levels, also the Reynolds shear stress (see inset of Fig. \ref{fig:MDR_friction}) smoothly links low Reynolds number EIT and high Reynolds number MDR whereas Newtonian turbulence levels are an order of magnitude larger \cite{warholic_influence_1999}. The same holds for velocity fluctuations (see also Fig. \ref{fig:far_onset}(a)). It should also be taken into account that Newtonian turbulence necessarily arises via spatially localized structures (puffs and slugs) and spatiotemporal intermittency.  These localized structures require finite amplitude fluctuation and friction levels and hence do not smoothly develop from low levels. EIT on the other hand is never spatially localized but always space filling. A feature that persists during its development to high $Re$ and MDR.  Spatio temporal intermittency,  characteristic for the transition to Newtonian turbulence, is absent.

	In summary, we have shown in experiments  that EIT arises from a center mode predicted by linear stability analysis \cite{garg2018}. From theoretical considerations it is evident that this mode requires finite inertia \cite{garg2018} and hence the EIT instability indeed requires both inertia and elasticity. This observation rules out a direct connection between EIT and purely elastic turbulence \cite{groisman2000elastic}. On the other hand the transition is considerably more complex than the instability suggested by linear analysis \cite{garg2018}. Although fluctuation amplitudes appear to increase continuously and at first sight seem to support a linear instability and a super-critical scenario, the onset of EIT can be pushed to Reynolds numbers more than an order of magnitude lower than permitted by the linear theory. Moreover the chaotic three dimensional motion detected even close to onset testifies that nonlinear effects must be taken into account. Both these observations are indicative for a sub-critical scenario. With increasing $Re$ fluctuation amplitudes eventually strongly increase when the centre mode is replaced by a wall mode consisting of inclined streaks and strong shear layers. This structural change occurs at $Re$ of order $10^2$ and hence far below the inertia levels required for Newtonian type turbulence. The resulting flow pattern  remains qualitatively unchanged with increasing $Re$ demonstrating that EIT is active in the maximum drag reduction limit and hence inhibits flow laminarisation even if polymers ultimately were to completely eradicate ordinary turbulence.

		{\bf Methods}
		
		Experiments are carried out in a $1.1$~m long smooth glass pipe with an inner diameter $D=4$~mm.
		A smooth inlet ensures that the gravity driven water flow remains laminar to $Re$ greater than $5000$. Starting from $150D$ downstream of the inlet, the pressure drop is measured over a pipe length of $75D$ using a differential pressure sensor (DP$15$ -- Validyne Engineering). Directly downstream, an identical sensor is used to measure pressure fluctuations over a pipe length of $4D$. A planar particle image velocimetry (PIV) system (LaVision GmbH), located $250D$ downstream of the pipe inlet, is employed to monitor the velocity field in a radial-axial cross section. At the same location and positioned at the pipe center, a Laser Doppler velocimetry (LDV) system (Powersight -- TSI GmbH) is used to measure the axial velocity component.
		
		The working fluid is a $600$ ppm (parts per million by weight) PAAm (Polyacrylamide with molecular weight of $5\times 10^6$~Da, Lot\#685910 -- Polysciences, Inc.) solution in either water or water glycerol mixtures ($50\%$ and $66\%$ glycerol). The addition of glycerol effectively increases the viscosity of the Newtonian solvent and allows us to investigate flows at low Reynolds numbers while keeping the shear rates and hence elastic forces (or more precisely the Weissenberg number) high. The chosen polymer concentration approaches the upper end of the dilute limit (estimated from the measure of intrinsic viscosity to be $\approx700$~ppm).
		

\onecolumngrid
\clearpage
\begin{center}
	\textbf{\large Supplemental Information: Experimental observation of the origin and structure of elasto-inertial turbulence}
\end{center}
\setcounter{equation}{0}
\setcounter{figure}{0}
\setcounter{table}{0}
\setcounter{page}{1}
\makeatletter
\renewcommand{\theequation}{S\arabic{equation}}
\renewcommand{\thefigure}{S\arabic{figure}}
\renewcommand{\bibnumfmt}[1]{[S#1]}
\renewcommand{\citenumfont}[1]{S#1}

\section*{Linear stability analysis: equations and methodology}

We consider the motion of an incompressible viscoelastic fluid
flowing through a pipe of uniform circular section. The dynamics in
this problem is governed by the continuity and Navier-Stokes equations,
along with a constitutive equation to model polymer dynamics. The latter
equation describes the temporal evolution of a polymer conformation tensor,
$\mathbf{C}$, that contains the ensemble average elongation and orientation
of all polymer molecules in the flow. A simple Hookean dumbbell model
is used to represent the polymer molecules. Normalizing velocity and
length with the laminar centreline velocity $u_c$ and the pipe radius $R$,
the pressure with the dynamic pressure, $\rho u_c^2$,
where $\rho$ is the fluid's density, and the polymer conformation tensor with $kT/H$,
where k denotes the Boltzmann constant, $T$ is the absolute temperature and $H$ is
the spring constant, the dimensionless equations read
\begin{equation}\label{eq:gov_eq}
		\nabla \cdot \mathbf{v} = 0,\\
		\partial_t\mathbf{\mathbf{v}} + \mathbf{v}\cdot\nabla\mathbf{v} = 
		-\nabla P + \frac{\beta}{Re} \nabla^2\mathbf{v} + \frac{(1-\beta)}{Re} \nabla\cdot \mathbf{T}
		+ \frac{4(1+\alpha)}{Re}\hat{e_z},\\
		\partial_t\mathbf{C}+\mathbf{v}\cdot\nabla\mathbf{C} =
		\mathbf{C}\cdot\nabla\mathbf{v} + (\nabla\mathbf{v})^T\cdot\mathbf{C}- \mathbf{T},	
\end{equation}
where $\mathbf{v}=(u,v,w) $ is the velocity vector field in cylindrical coordinates $(z,r,\theta)$, $Re= u_c R/\nu$ is the Reynolds number
and $\alpha$ is the fluctuating pressure gradient required to impose a constant flow rate. Polymers are coupled to the Navier-Stokes equations
through the polymer stress tensor $\mathbf{T}$, which is calculated using the Oldroyd-B model~\cite{oldroyd1950},
\begin{equation}\label{eq:OlroydB}
	\mathbf{T} = \frac{1}{Wi}(\mathbf{C}-\mathbf{I}),
\end{equation}
where $\mathbf{I}$ is the unit tensor and $Wi$ is the Weissenberg number; a dimensionless number measuring
the ratio of the polymer relaxation time $\lambda$ to the characteristic flow time scale  $R/u_{c}$.
Despite the Oldroyd-B model relies on the unrealistic assumption that polymers have infinite extensibility,
it has been shown to be a good model of highly elastic polymers, i.e. Boger like fluids,
such as those considered in this study.

Equations  \eqref{eq:gov_eq} admit an analytical solution for the steady laminar flow. As in the Newtonian case, the laminar velocity is the classical Hagen-Poiseuille flow, $U = 1- r^2$. The nonzero components of the symmetric polymer conformation tensor under base flow conditions
are $C_{zz}=4r^2Wi^2+1$, $C_{rz}=-2rWi$,$C_{rr}=1$ and $C_{\theta\theta}=1$. Equations~\eqref{eq:gov_eq} were linearized around the analytical base flow to investigate its linear stability. The fluctuating velocity fields, pressure and polymer conformation
tensor were expanded in Fourier series in the axial and azimuthal directions, whereas eighth order finite differences on a Gauss-Lobatto-Chebyshev grid were used to discretize the radial derivatives. The largest eigenvalues dictating the stability of the base flow were determined by time integrating the linearized
equations. The simulations were initialized with disturbances of small amplitude satisfying the boundary conditions, i.e no slip at the wall and zero divergence, and the temporal evolution of the amplitude of these disturbances was monitored. After an initial transient, the
amplitude may exhibit an exponential growth (decay) if the flow is unstable (stable) or it may remain constant if the flow is neutrally stable, i.e. under critical conditions. In the latter case,
the corresponding eigenvalue is zero, whereas in the other cases the leading eigenvalues are easily obtained by measuring the growth or decay rates. Note that since nonlinearity is absent in these
simulations, a saturated state is never reached and the kinetic energy keeps growing or decaying at a constant rate as time is evolved in the simulation. This methodology to calculate the leading eigenvalues is equivalent to the widely used power method. Our code was validated by reproducing several results published in the literature for both the Newtonian and Non-Newtonian cases. Some examples are illustrated next. 

\begin{figure}
	\begin{tabular}{cc}
		$(a)$ & $(b)$\\
		\includegraphics[width=0.3\linewidth]{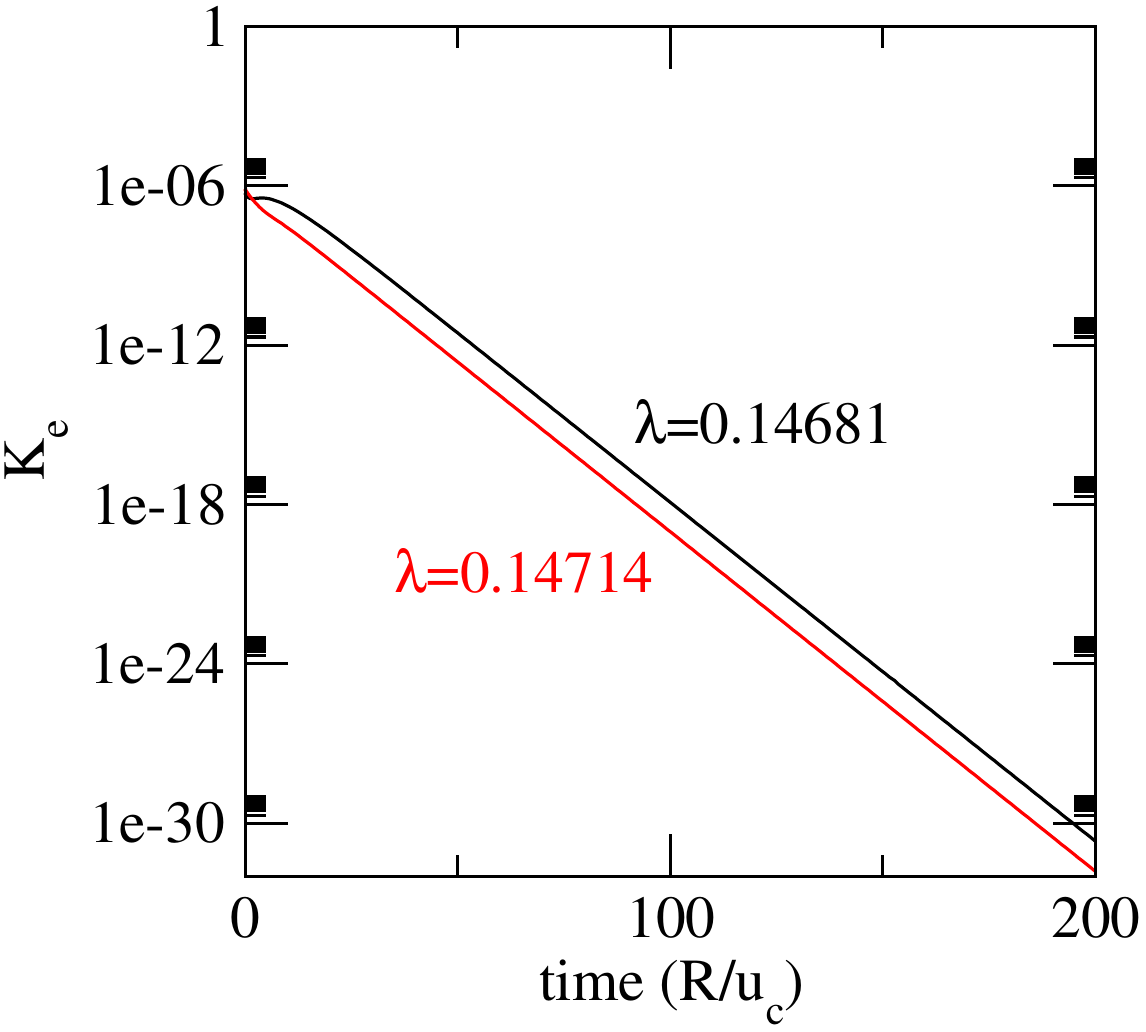} &
		\includegraphics[width=0.3\linewidth]{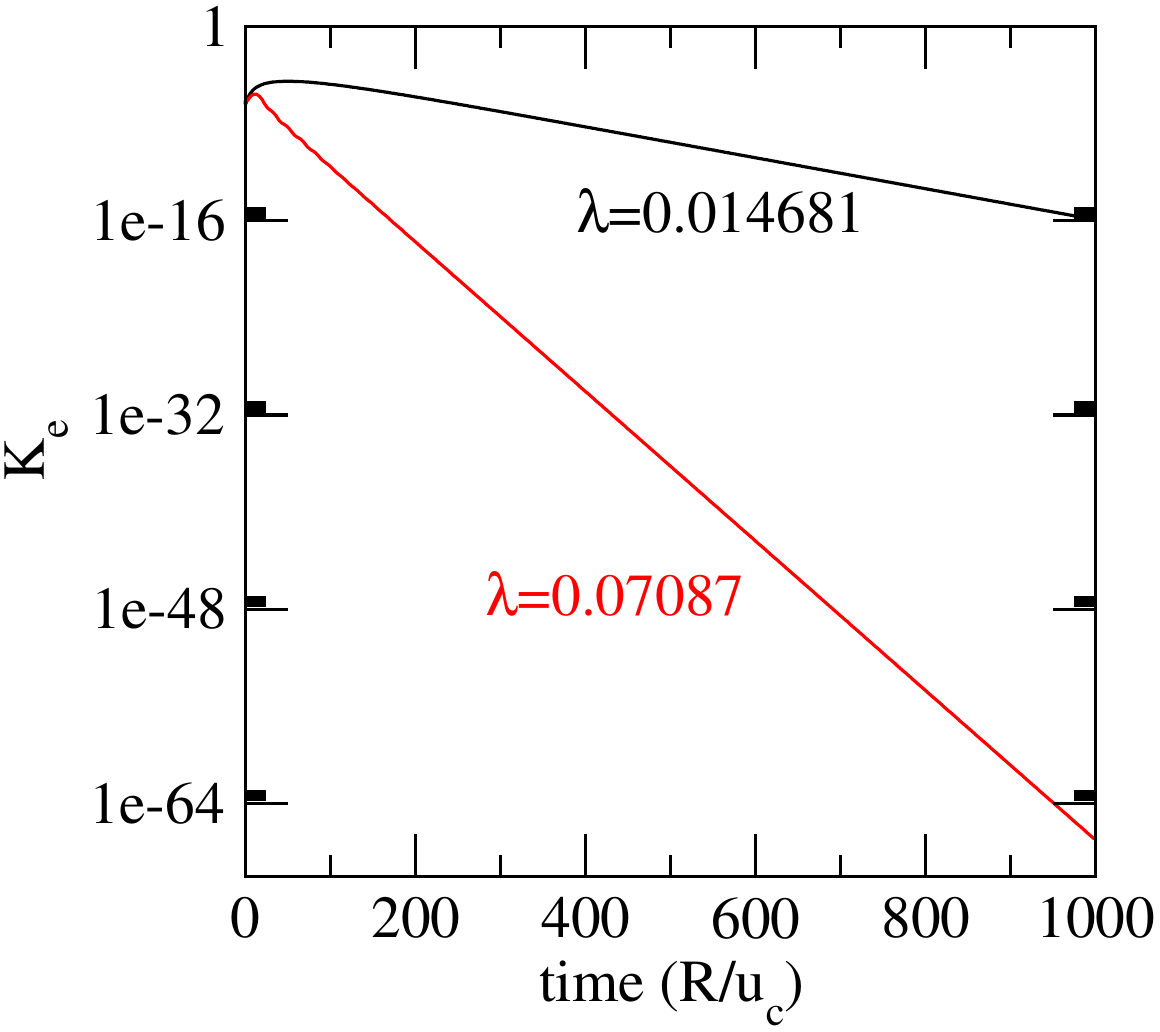}\\
	\end{tabular}
	\caption{Decay of the kinetic energy of the Fourier modes (n,l) = (1,0) and (1,1)
		in simulations at $Re =100$ $(a)$ and $1000$ $(b)$. Here, $n$ and $l$ indicate azimuthal
		and axial Fourier modes respectively. The axial wavenumber in these simulations
		is $k_z=1$, i.e. the pipe length is $2\pi$. The leading eigenvalues $\lambda$
		are obtained from exponential fits of these data.}\label{fig:eigenvalues_Newt}
\end{figure}

Figure~\ref{fig:eigenvalues_Newt} shows the temporal evolution
of the kinetic energy corresponding to the Fourier modes (n,l) = (1,0) and (1,1) in Newtonian pipe flow for
$Re=100$  and $Re=1000$ in simulations performed with $k_z=1$. The eigenvalues estimated from
the decay rates (indicated in the figure) are in excellent agreement with those reported in~\cite{meseguer2003}
where a formal stability analysis was carried out using a Petrov-Galerkin formulation.

\begin{figure}
	\begin{tabular}{cc}
		$(a)$ & $(b)$\\
		\includegraphics[width=0.3\linewidth]{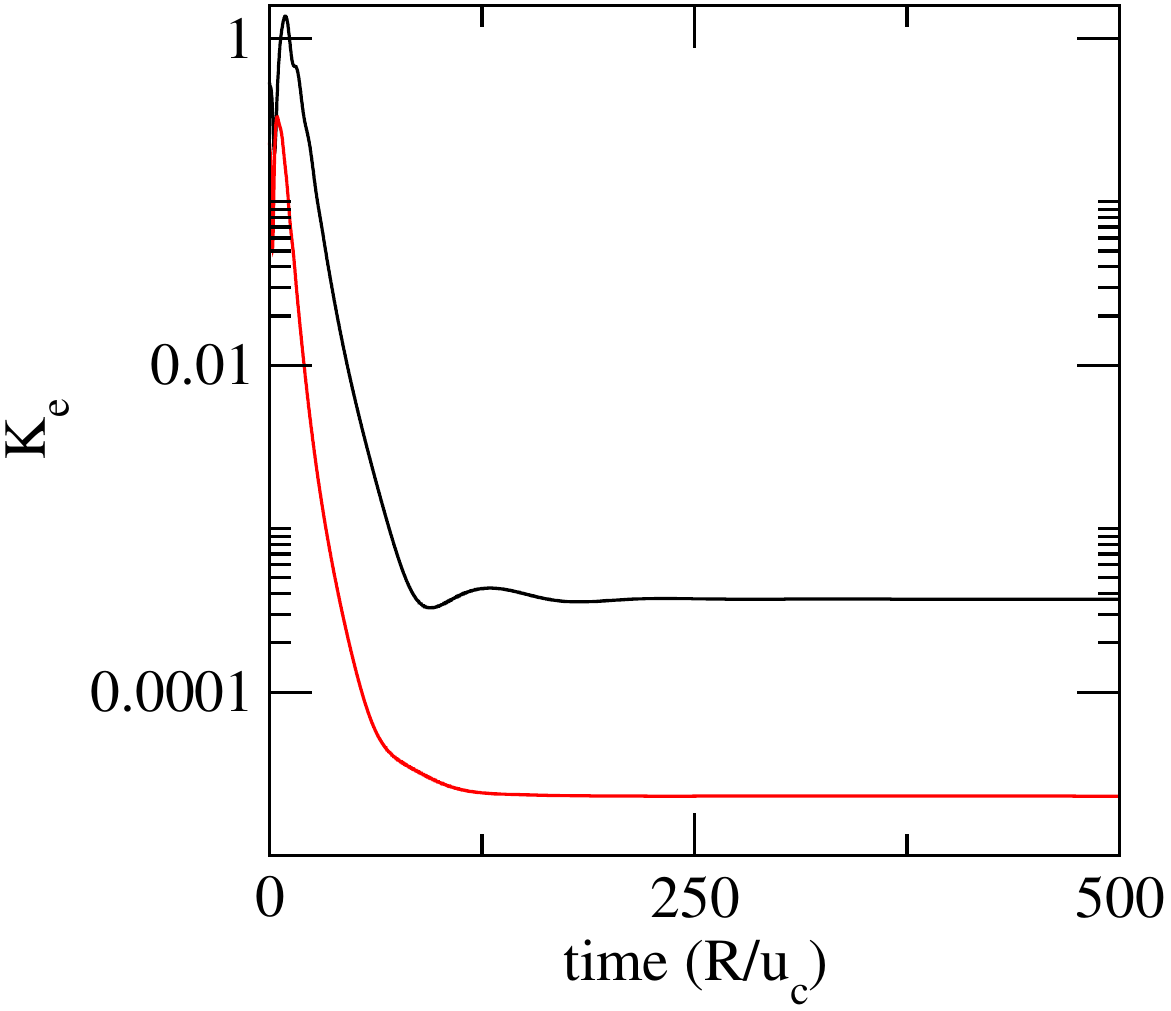} &
		\includegraphics[width=0.3\linewidth]{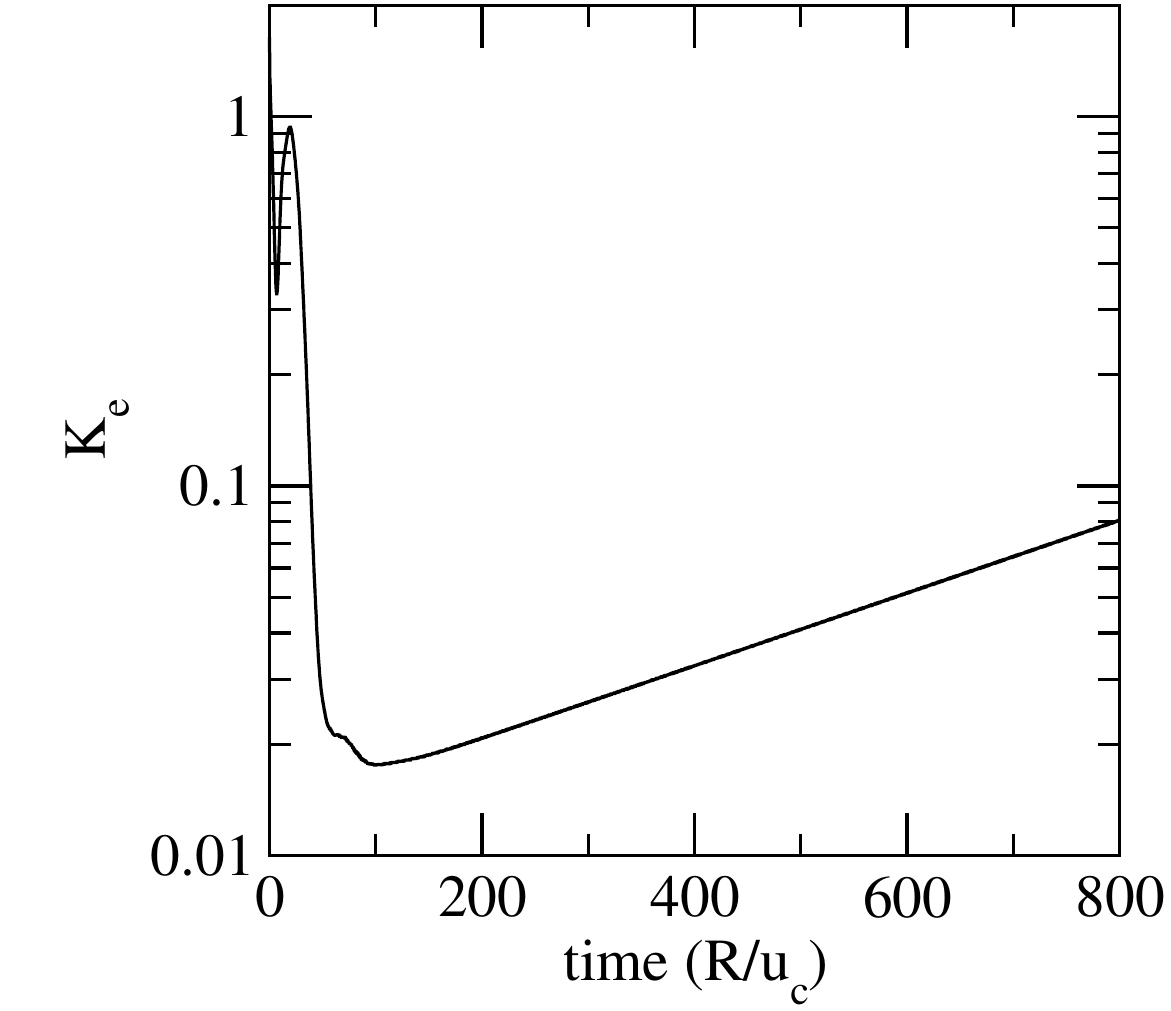}\\
	\end{tabular}
	\caption{ $(a)$ Kinetic energy for critical disturbances. The black line corresponds to a simulation with
		$\beta=0.25$, $Re=5026.14$, $Wi=50.26$ and $k_z=2.80$, whereas the red line was calculated at
		$\beta=0.5464$, $Re=6871.27$, $Wi=68.71$ and $k_z=5.15$. $(b)$ Growth of the kinetic energy for an
		unstable case calculated at $\beta=0.65$, $Re=800$, $Wi=65$ and $k_z=1$.}\label{fig:critical_and_unstable}
\end{figure}

\begin{figure}
	\begin{tabular}{cc}
		$(a)$ & $(b)$\\
		\includegraphics[width=0.3\linewidth]{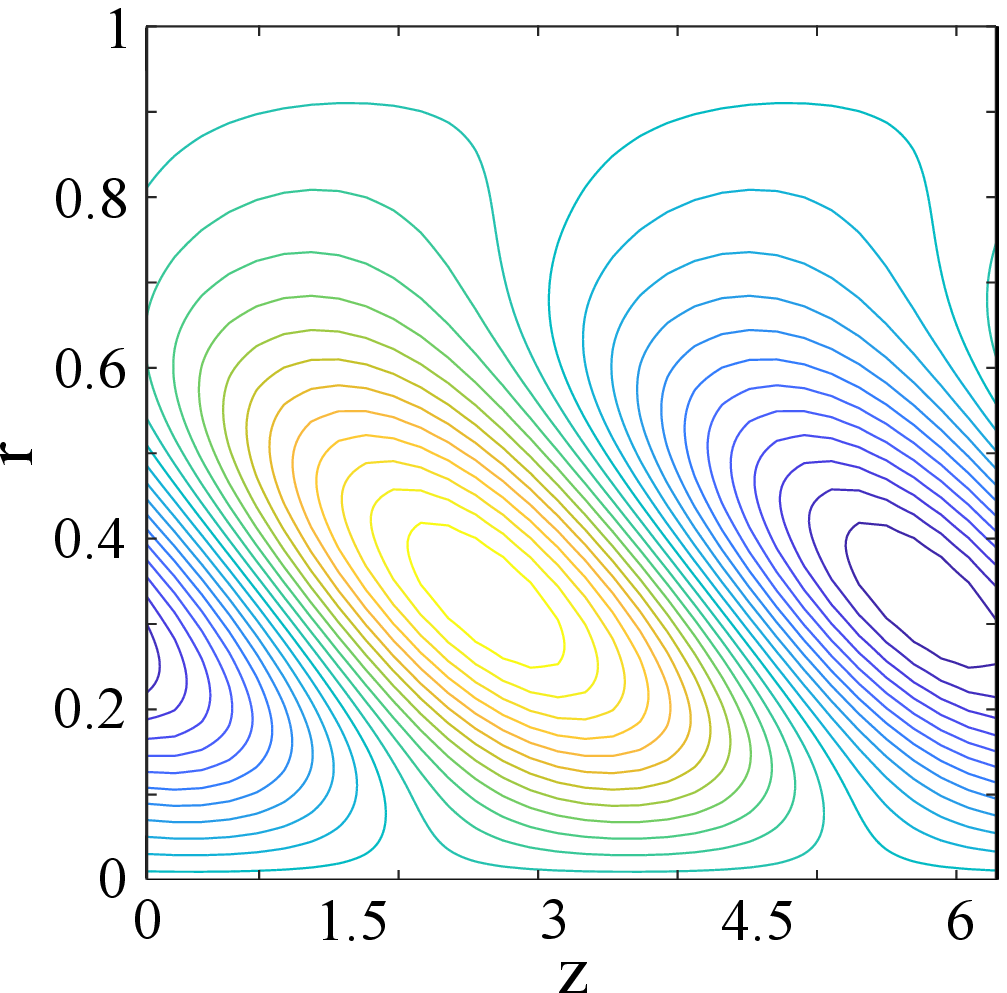} &
		\includegraphics[width=0.3\linewidth]{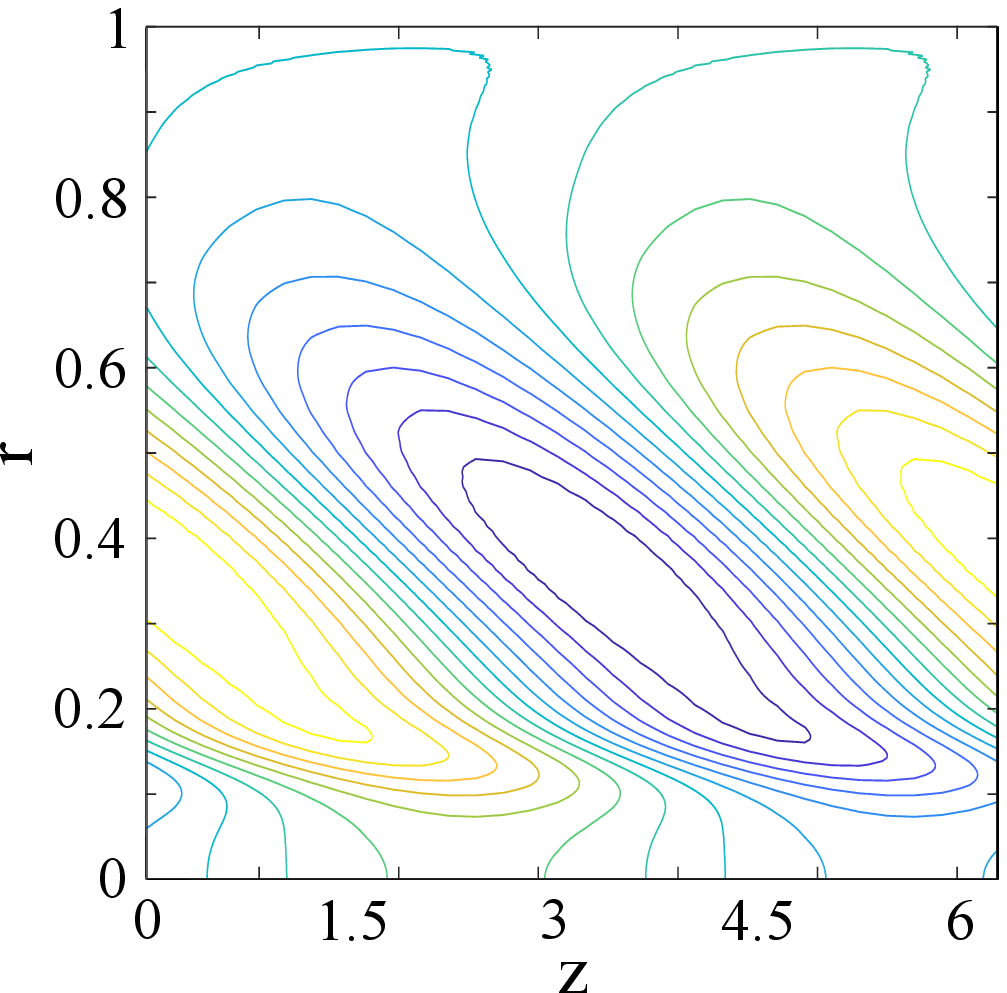}\\
	\end{tabular}
	\caption{Contour plots of the radial velocity disturbance $(a)$ and the polymer force ($\nabla \cdot T$) $(b)$
		for the unstable mode at $\beta=0.65$, $Re=800$,$Wi=65$ and $k_z=1$.}\label{fig:unstable_mode}
\end{figure}

To validate the viscoelastic code, several critical cases reported in~\cite{garg2018} were successfully reproduced.
Fig~\ref{fig:critical_and_unstable} (a) shows the kinetic energy for two combinations of critical parameters taken from Figure 5 in~\cite{garg2018}.
As seen, after the initial transient the kinetic energy neither grow nor decay, confirming that the flow is neutrally
stable for these values of the control parameters. Finally, the growth of the kinetic energy for an unstable case calculated
at $Wi=50$, $\beta=0.5$,$k_z=1$ and $Re=800$ is shown in Fig.~\ref{fig:critical_and_unstable} $(b)$.
Contour plots of the radial velocity fluctuation and the polymer force of the unstable mode
are illustrated in figures~\ref{fig:unstable_mode} $(a)$ and $(b)$. These plots replicate the figure 2
in~\cite{garg2018}, which was calculated for the same values of the control parameters following a different
methodology.

\end{document}